\documentclass[preprint,12pt]{elsarticle}
\journal{Journal of Computational Physics}

\usepackage{amsmath}
\usepackage{amssymb}
\usepackage{amsthm}
\DeclareMathOperator\erf{erf}

\begin{document}

\begin{frontmatter}

\author{Samuel Temple Reeve}
\author{Alejandro Strachan \corref{cor2}}
\cortext[cor2]{Corresponding author.}
\ead{strachan@purdue.edu}

\address{School of Materials Engineering and Birck Nanotechnology Center, 
Purdue University, West Lafayette, IN 47906, USA}

\title{Error correction in multi-fidelity molecular dynamics simulations using functional uncertainty quantification}

\begin{abstract}
We use functional, Fr\'{e}chet, derivatives to quantify how thermodynamic outputs of a molecular dynamics (MD) simulation depend on the potential used to compute atomic interactions. Our approach quantifies the sensitivity of the quantities of interest with respect to the input \textit{functions} as opposed to its parameters as is done in typical uncertainty quantification methods. We show that the functional sensitivity of the average potential energy and pressure in isothermal, isochoric MD simulations using  Lennard-Jones two-body interactions can be used to accurately predict those properties for other interatomic potentials (with different functional forms) without re-running the simulations. This is demonstrated under three different thermodynamic conditions, namely a crystal at room temperature, a liquid at ambient pressure, and a high pressure liquid. The method provides accurate predictions as long as the change in potential can be reasonably described to first order and does not significantly affect the region in phase space explored by the simulation. The functional uncertainty quantification approach can be used to estimate the uncertainties associated with constitutive models used in the simulation and to correct predictions if a more accurate representation becomes available.  
\end{abstract}


\end{frontmatter}

\section{Introduction}
Uncertainty quantification (UQ) is becoming increasingly important in predictive simulations of materials and devices \cite{Chernatynskiy2013}. While the majority of the early UQ work focused on solid and fluid mechanics, there is growing interest in applying and extending UQ techniques to simulations at the material level, including density functional theory \cite{Mortensen2005, Anderson2012}, molecular dynamics (MD) \cite{Rizzi2012p1, Rizzi2013p1, Angelikopoulos2013, Karniadakis2015, Patrone2016}, and multiscale methods \cite{Koslowski2011, Kim2013, Vedula2013}. These efforts highlighted the importance of acknowledging uncertainties in model parameters, from measurement or averaging techniques, as well as intrinsic variability of the systems or processes under investigation. These studies, and most UQ work thus far, examined uncertainties in input parameters and specialized software packages have been developed for this task: examples include DAKOTA \cite{Adams2014}, PUQ \cite{Hunt2015}, and \(\Pi\)4U \cite{Hadjidoukas2015}. However, in many applications -- especially those involving complex physics -- accounting for uncertainty in parameters is not enough, as the functional forms of the constitutive models themselves are approximate. This is particularly true in materials modeling where lack of knowledge leads to unquantified or poorly quantified uncertainties. Examples of input functions with varying degrees of accuracy include exchange and correlation functionals used in density functional theory calculations \cite{Mortensen2005, Aldegunde2016}, interatomic potentials for molecular dynamics (MD) \cite{Rizzi2012p2, Rizzi2013p2, Farrell2015}, generalized stacking faults used in dislocation dynamics \cite{Cao2015}, and constitutive laws for micromechanical simulations. In this paper we use functional derivatives (FD), recently proposed as a mathematical framework to quantify uncertainties that arise from constitutive models used in simulations \cite{Strachan2013},  to quantify and correct uncertainties that originate from the interatomic potential used in MD simulations. 

Functional uncertainty quantification (FunUQ) can, in principle, be used to assess the uncertainties originating from approximate input constitutive laws, correct predictions if a more accurate function becomes available, and rank when and where to replace a low-fidelity model used in a simulation with one of higher fidelity in order to reduce prediction error by running additional simulations. This paper introduces a computationally efficient method to compute FDs in MD simulations involving two-body interatomic potentials, extending ideas from thermodynamic integration and free-energy perturbation methods \cite{Frenkel2001, Chipot2007}.  The FD quantifies how a quantity of interest (QoI) -- in this case the total potential energy or pressure computed from an MD simulation in the canonical ensemble -- depends on the input function, here the Lennard-Jones (LJ) two-body pair potential. We further show that the FD with respect to the LJ potential can be used to compute accurate correction to the potential energy and pressure for a family of pair potentials without re-running the simulation. This is true as long as the discrepancy between the potentials remains within reasonable bounds and the phase space explored by the system with the new potential is not significantly different from that of the original. 

The remainder of the paper is as follows. Section 2 describes the functional approach to UQ and connections to similar UQ methods, followed by simulation details and methods to calculate the functional derivative numerically in Section 3. In Section 4 we describe specific examples of error correction using FunUQ with low and high-fidelity models. Section 5 discusses results and concludes the paper.

\section{Functional uncertainty quantification} 
In general, a simulation predicts a quantity of interest, \(Q\), given some set of input parameters \(P_i\) and constitutive functions \(f_j\), themselves functions of an independent variable \(z\) and input parameters \(N_k\):
\begin{equation} \label{eq:QoI}
Q=Q({P_i},{f_j(z,{N_k})}) 
\end{equation}
In the cases of interest here, \(Q\) will be the time averaged potential energy or pressure and \(f(z)\) the LJ potential as a function of interatomic distance used in the MD simulation. 

The problem of forward propagation in UQ is most commonly concerned with uncertainty in simulation outputs arising from uncertainty in the input parameters. This approach, while very valuable, ignores the fact that the functional forms used as constitutive laws are, almost invariably, approximate and lead to errors in the simulation. To quantify uncertainties with respect to input functions we utilize functional derivatives (Fr\'{e}chet derivatives) of the QoI with respect to the input functions. The FD can be written in differential form as:
\begin{equation} \label{eq:FD1}
\frac{\delta Q[f]}{\delta f(z)}(z_i) = \lim_{\epsilon \to 0} \frac{Q[f(z) + \epsilon\cdot\delta(z-z_i)] - Q[f(z)]}{\epsilon}
\end{equation}
This characterizes the functional sensitivity of the QoI with respect to the input function. The definition in Eq. \ref{eq:FD1} uses the Dirac delta function as the functional variation; as we describe below, to calculate the functional derivative numerically for the MD simulations we use narrow Gaussian distributions centered at \(z_i\).

Besides quantifying uncertainties in the prediction given uncertain input functions, the functional derivative can be used to correct the error that arises from the use of a low-fidelity model \((f)\) if one of higher-fidelity \((g)\) becomes available. A first order correction is then obtained using the product of the functional sensitivity (the FD) and functional discrepancy \((g(z)-f(z))\). We call this product the functional error and the first order correction for the QoI is:
\begin{equation} \label{eq:correct}
\Delta Q = \int \frac{\delta Q[f]}{\delta f(z)} \cdot (g(z)-f(z)) dz 
\end{equation}
This is an extension of the multi-variate UQ expression to the space of input functions: sum of variables is replaced by an integral, derivatives with respect to individual variables are replaced by the functional derivative, and the functional discrepancy takes the place of the difference in the values of input variable. A simple analytical example is included the supplementary material, Section S1.

In the absence of a high-fidelity model Eq. \ref{eq:correct} can alternatively be used for uncertainty propagation by replacing the discrepancy with the uncertainty in \(f(z)\) and taking the absolute value of the functional derivative \cite{Strachan2013}. This procedure returns a first order bound of the uncertainty in Q. 

Additionally, this equation can be used to rank high-fidelity simulations in order of their functional error to optimally reduce the error in the predicted QoI, recently demonstrated with calculation of the functional derivative for the restoring force in a multi-fidelity radio frequency MEMS switch simulation \cite{Strachan2013}. This functional derivative was used to rank model evaluations and minimize the necessary computational cost to maximize error correction in the simulation. In Section 3 we extend the formulation to a significantly more challenging problem: molecular dynamics simulations. 

Many other approaches in utilizing and optimizing multi-fidelity simulations and describing model discrepancy exist, notably stemming from the work of Kennedy and O'Hagan\cite{Kennedy2001}. Multi-fidelity simulation orchestration with Bayesian approaches optimize (with uncertainty) utilization of multiple levels of model fidelity\cite{Kennedy2000} and include uncertainty from model selection within a simulation framework, most often through model averaging \cite{Zhang2000, Droguett2008, Park2010}. Model discrepancy approaches most often investigate the difference between the simulation predictions and observed data (structural uncertainty)
\cite{Strong2014}. Alternatively, the discrepancy is between the computational model and a surrogate is of interest\cite{Leary2004}. In contrast, we describe direct consideration and correction of uncertainty from multiple models with varying fidelity.

\section{Functional derivatives in molecular dynamics}
\subsection{Systems of interest and simulation details}

We demonstrate the FunUQ method in molecular dynamics simulations where the input function is a pairwise interatomic potential and the QoIs are the potential energy and pressure (long-time averages) of the system. MD simulations use this pairwise potential energy as a function of atomic separation as an input to compute total energy and interatomic forces (obtained as the negative gradient) which are used to integrate Newton's equations of motion to predict the time evolution of the system. We take the Lennard-Jones 12-6 potential as the low-fidelity input function: 
\begin{equation} \label{eq:LJ}
\phi_0(r) = 4\epsilon((\sigma/r)^{12}-(\sigma/r)^6) 
\end{equation}
and test the ability of FunUQ to correct the prediction of the QoI for a family of high-fidelity potentials. The LJ potential uses an inverse sixth order term to describe the attractive part of the interactions and an inverse twelfth order term to describe shorter range repulsion. The  parameters for the low-fidelity LJ potential are designed to roughly describe copper: the equilibrium bond distance, \(\sigma=2.315\)\AA\ and the equilibrium energy well depth, \(\epsilon=0.167 eV\) \cite{Wolf1992}. These values were fit to the bulk melt temperature and room temperature lattice constant. We note that this is not an accurate potential for Cu as it ignores important many body effects critical to describe elastic constants and defect energetics. Also notable is the discrepancy of the liquid densities as compared to experiment \cite{Cahill1962}.  However, the goal of the paper is to demonstrate the applicability of FunUQ to an MD problem. In this spirit, the high-fidelity potentials are similarly designed for demonstration purposes only and do not represent a more accurate representation of an actual material. We construct seven pair potentials by additively modifying the LJ potential with sine functions in Table 1. This family of functions will be denoted Sine 1 to Sine 7.

\begin{table}
  \centering
  \caption{High-fidelity potential sine modification terms}
  \label{tbl:sine}
  \begin{tabular}{ll}
    \hline
    Name  & Modification function  \\
    \hline
    Sine 1 & \(0.44+0.46\sin(0.17(24.2+r))\)   \\
    Sine 2 & \(-0.47\sin(-0.15(14+r))\cdot \exp(-r)\)  \\
    Sine 3 & \(0.07\sin(1.2(-1.2+r))/r^2\)  \\
    Sine 4 & \(-0.01+0.2\sin(0.3(14+r))\cdot \exp(-r)\)  \\
    Sine 5 & \(0.7\sin(0.4(11+r))\cdot \exp(-r)\)  \\
    Sine 6 & \(0.9\sin(0.4(11+r))\cdot \exp(-r)\)  \\
    Sine 7 & \(1.1\sin(0.4(11+r))\cdot \exp(-r)\)  \\
    \hline
  \end{tabular}
\end{table}
A final high-fidelity model tested is the Morse potential, defined as the sum of two exponentials:
\begin{equation} \label{eq:Morse}
\phi (r) = D_0(e^{-2\alpha(r-r_0)} -2e^{-\alpha(r-r_0)}) 
\end{equation}
We use \(D_0=0.161 eV\), \(\alpha=2.09\)\AA\({}^{-1}\), and \(r_0=2.62\)\AA\ to similarly roughly describe copper. 

\begin{figure} 
  \makebox[\textwidth][c]{\includegraphics[width=3.3in]{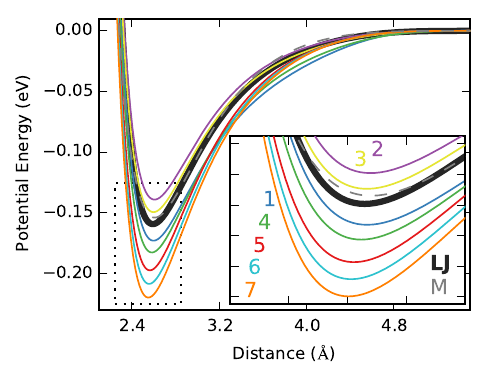}}
  \caption{Interatomic potentials used with inset taken from the dotted region. High-fidelity models obtained as sine modified LJ potentials are shown in color labeled with numbers, low-fidelity LJ with bold line in black, and Morse (M) with dashed line in gray.}
  \label{fig:potentials}
\end{figure}

For all potentials a smoothing function is used to ensure stable dynamics near the cutoff. The function is of fourth order to create potential energy and force curves that both smoothly tend to zero:
\begin{equation} \label{eq:smooth}
s(r) = \frac{(\frac{(r - r_c)}{w})^4}{(1+\frac{(r - r_c)}{w})^4}
\end{equation}
where \(r_c\) is the cutoff distance and \(w\) is the width of the smoothing. For all potentials the cutoff is 5.79\AA\ (\(2.5\sigma\), commonly used for LJ potentials) and smoothing width 1.5\AA. Each potential curve is created by summing the base equation and the modification term and subsequently taking the product with the smoothing function. The force curve is then created with product rule differentiation of the potential. These potentials are shown in Figure \ref{fig:potentials}. 

Each system consists of 500 atoms simulated under isothermal, isochoric conditions (canonical ensemble). Three physical states of the system were simulated: one solid at ambient temperature and pressure (\(300 K\) with density of \(9.02 g/cm^3\)), one liquid slightly above the melt temperature at ambient pressure (\(1300 K\) with density of \(6.48 g/cm^3\)) and another liquid at extreme temperature and pressure (\(5000 K\) with density of \(8.93 g/cm^3\), corresponding to \(55 GPa\)). 

All MD simulations were performed using the LAMMPS package \cite{Plimpton1995}. In order to obtain the quantities of interest, energies, pressures and functional derivatives, we performed both temporal and ensemble averages as is common practice in MD simulations. For each condition, we used several simulations starting with statistically independent velocities obtained from the Maxwell-Boltzmann distribution at the desired temperature. For each high-fidelity potential and thermodynamic condition of interest we performed 16 independent simulations, each 1 ns long, sufficient for good statistical sampling of the quantities of interest. For each low-fidelity potential and thermodynamic condition, the total sample time was 64 ns, from 64 independent simulations of 1 ns. A longer simulation time was necessary for the low-fidelity simulations in order to properly converge the FD with the perturbative approach (discussed in the following section). The use of multiple independent simulations allowed for concurrent computation and reduced wall-clock time for the study. An example of the convergence of the FunUQ corrections is shown in the supplementary material, Section S2. We now consider the calculation of the FD for MD simulations.

\subsection{Numerical functional derivatives and a perturbative approach for their calculation}

For the specific case of interest the general functional derivative expression from Eq. \ref{eq:FD1} becomes:
\begin{equation}  \label{eq:FD2}
\frac{\delta Q[\phi]}{\delta \phi}(r_i) = \lim_{\epsilon \to 0}{\frac{Q[\phi_{0}(r) + \epsilon\cdot\phi'(r-r_i)] - Q[\phi_{0}(r)]}{\epsilon} 
}\end{equation}
where \(Q[\phi]\) denotes the average QoI -- the potential energy or pressure of the system -- using interatomic potential \(\phi\). The unmodified LJ potential is \(\phi_0\) and \(\phi'\) is a normalized Gaussian perturbation centered at interatomic distance \(r=r_i\) with width \(\sigma\). Note that \(\phi_0\) here is the same low-fidelity function used for the correction in Section \ref{sec:results} (there denoted \(\phi_{LF}\)) .

\begin{figure} 
  \makebox[\textwidth][c]{\includegraphics[width=3.3in]{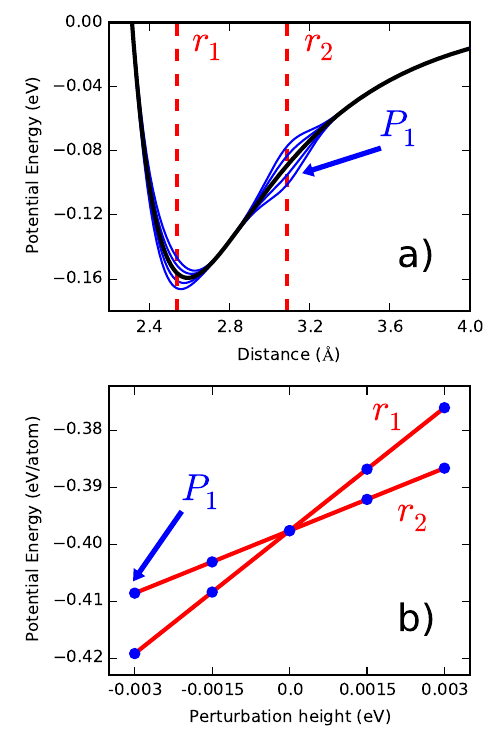}}
  \caption{Numerical calculation of a functional derivative with a) unmodified LJ potential in bold black and eight examples of modified interatomic potentials in blue (two positive and two negative height Gaussian perturbations with fixed width at each of two positions, \(r_1\) and \(r_2\)) and b) numerical derivatives with respect to perturbation height from the example perturbations in a). Perturbations enlarged for visibility with one example, \(P_1\), marked in both a) and b).}
  \label{fig:calcFD}
\end{figure}

The functional derivative in Eq. \ref{eq:FD2} can be calculated by performing a set of simulations with the low-fidelity potential modified by perturbations of varying \(\epsilon\) and computing a numerical derivative for each position \(r_i\). These modified Lennard Jones potentials and resulting average potential energies as a function of size of the perturbation \(\epsilon\) are shown in Fig. \ref{fig:calcFD}(a) and (b), respectively (discussed in more detail below). However, this brute force approach is computationally very intensive considering that we may need to sample the functional derivative at 100 values of \(r_i\). Even with only three values of \(\epsilon\) for each separation distance one would need to perform 300 separate MD simulations. To alleviate the computational cost of the approach we now derive a perturbative approach to calculate the functional derivative that can be computed with little overhead with respect to the nominal simulation using the low-fidelity potential, \(\phi_0\).

At the heart of the calculation of the FD is the evaluation of canonical averages of the QoIs with the interatomic potential and a Gaussian perturbation. Recognizing that the modified Hamiltonian of the system can be additively decomposed into the kinetic energy, the original LJ potential energy and the functional variation (or perturbation) for these two-body potentials we can write the canonical ensemble average of quantity Q as:
\begin{equation} \label{eq:ensavg}
\langle Q \rangle_H = \frac{\int Q \cdot e^{-\beta H_0} \cdot e^{-\beta H'}}{\int e^{-\beta H_0} \cdot e^{-\beta H'}}
\end{equation}
where \(H_0\) is the Hamiltonian with the LJ potential and \(H'\) the potential energy resulting from the Gaussian perturbation following form from Eq. \ref{eq:FD2}. Equation \ref{eq:ensavg} can be re-written as the ratio between two canonical averages over the unmodified Hamiltonian by multiplying by additional factors of \(\int e^{-\beta H_0}\) and rearranging:
\begin{equation} \label{eq:perturbative}
\langle Q \rangle_H = \langle Q \cdot e^{-\beta H'} \rangle_{H_0} \cdot 
\frac{1}{\langle e^{-\beta H'} \rangle_{H_0}} 
\end{equation}
Since both canonical averages are over the unmodified potential only simulation with the low-fidelity potential is required. Therefore, computing the functional derivatives requires evaluating the Gaussian perturbation potentials on the trajectory obtained with the low-fidelity model. Such expressions are commonly used to compute free energies in thermodynamic integration and free energy perturbation approaches \cite{Straatsma1992, Kollman1993}.

Quantities for the canonical average in Eq. \ref{eq:perturbative} are computed every 1 ps with a total of 64 ns of simulation time with the low-fidelity LJ potential, deemed well converged (see supplementary material, Section S2). We evaluate these averages using a binned coordination number with a total of 2000 bins to compute \(H'\) and \(Q\)  (as it contains contributions from the perturbation), described in more detail in the appendix. By using the coordination number we further reduce computation and avoid modifying the MD code. Performing the calculation directly within the MD force loop with atomic positions would be equivalent excluding slight discrepancies from discretization. 

Using Eq. \ref{eq:FD2} with the first term in the difference calculated using Eq. \ref{eq:perturbative} we compute the functional derivative with respect to Gaussian perturbations centered at \(r_i\) with width \(\sigma = 0.1\) \AA\ and heights \(\epsilon=\pm0.00075\) and \(\pm0.00375\) eV (examples at two \(r_i\) in Fig. \ref{fig:calcFD}(a)). The perturbation width was chosen to minimize (localize) the perturbation while retaining sufficient sampling. The heights were similarly minimized, due to the difficulty of converging exponentially weighted averages, while ensuring a measurable effect from the perturbations. The numerical derivative is then evaluated by computing the slope of the QoI with respect to \(\epsilon\) (examples in Fig. \ref{fig:calcFD}(b)) at a set of interatomic separations ranging from zero past the potential cutoff in increments of \(0.05\) \AA.

\begin{figure} 
  \makebox[\textwidth][c]{\includegraphics[width=6.6in]{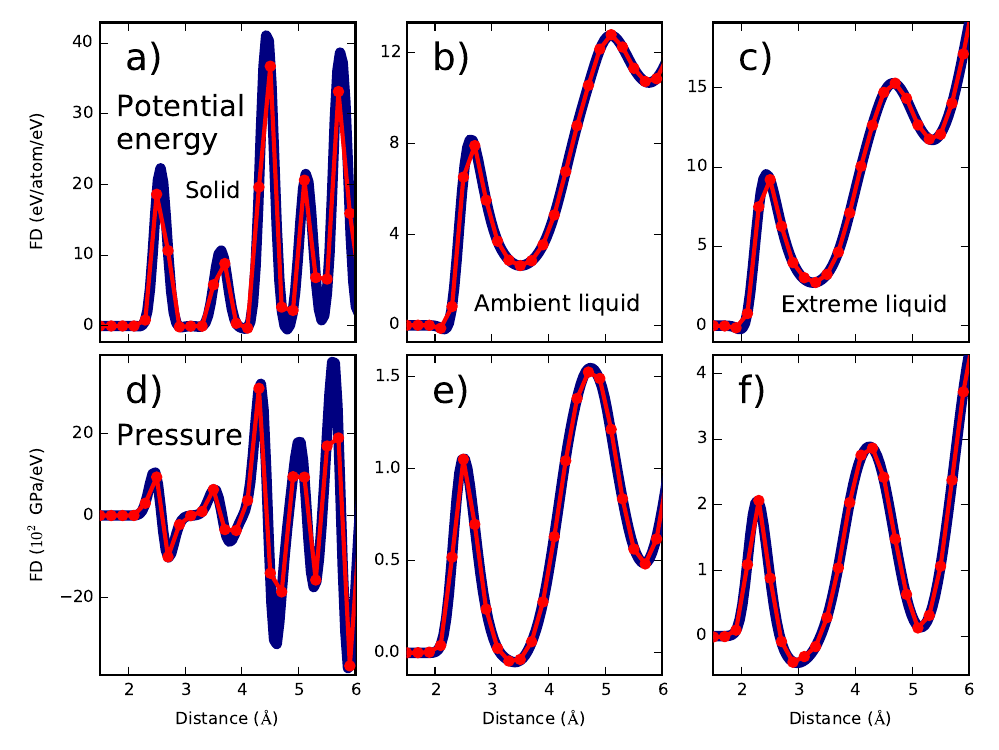}}
  \caption{Functional derivatives of the potential energy for the a) solid, b) ambient liquid, and c) extreme liquid cases and of pressure for the d) solid, e) ambient liquid, and f) extreme liquid cases. The perturbative method is shown in thick blue lines and the brute force method in red lines with points.}
  \label{fig:FD}
\end{figure}

The thick blue lines in Figure \ref{fig:FD} show the functional derivatives obtained in this manner for all physical conditions and both QoIs. These curves were averaged from multiple calculations of the FD with independent randomized samplings from the total of 64 ns of simulation time. The functional derivative displays significant information about the physics of the systems. The overall shapes correspond to the atomic shells as in the radial distribution function -- notably more pointed for the solid case. At very low separation distances the FD goes to zero as atoms hit the soft-wall repulsion of the potentials; for the extreme liquid system the atoms occupy smaller separations. 

To assess the accuracy of the calculation the functional derivative was also computed at equally spaced values of \(r_i\) using the brute force approach described above. For each value of \(r_i\) we perform four MD simulations with varying perturbations (identical to those in the perturbative approach) added to the LJ potential as shown in Fig. \ref{fig:calcFD}(a) and obtain the functional derivative with the same numerical derivatives as the perturbative approach. For each separate perturbation the simulation was run for 16 ns (split between 16 independent systems). The results of the brute force approach are shown in red in Fig. \ref{fig:FD}; the two methods of calculating the functional derivative are nearly identical with only small numerical discrepancies.

\section{Error correction using functional derivatives} \label{sec:results}
To demonstrate our approach we now use the functional derivative calculated perturbatively in the previous section to correct the potential energy and pressure predicted with the low-fidelity LJ potential assuming a more accurate function is available. As discussed in subsection 3.1, a family of seven high-fidelity potentials were created by modifying the LJ potential with sine functions; see Figure \ref{fig:potentials} and Table \ref{tbl:sine}. These results are discussed in sub-section \ref{sec:resultsSine}; the results for the Morse potential are discussed in sub-section \ref{sec:resultsMorse}.

\subsection{Corrections for sine modified potentials} \label{sec:resultsSine}
\begin{figure} 
  \makebox[\textwidth][c]{\includegraphics[width=3.3in]{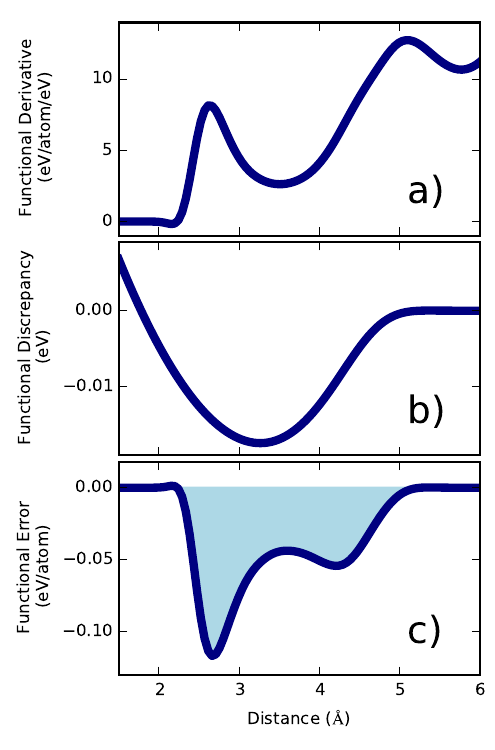}}
  \caption{Method of calculating the cumulative functional error with examples of the a) functional derivative , b) the functional discrepancy, and c) the product of the two for the ambient liquid with the Sine 1 high-fidelity potential. The shading shows the integrated area contributing to the total correction of the QoI as in Eq. \ref{eq:correct}.}
  \label{fig:correct}
\end{figure}

In order to use Eq. \ref{eq:correct} we need, in addition to the functional derivative, the discrepancy function \((\phi_{HF} (r) - \phi_{LF} (r))\),  the difference between the high and low-fidelity potentials. Figure \ref{fig:correct} illustrates a) the functional derivative, b) the functional discrepancy, and c) the product of the two, the functional error, each as a function of interatomic distance for one case, the ambient liquid with modified potential Sine 1. These results are shown for each physical case and potential in the supplementary material, Section S3. Note that the functional error goes to zero both for small and large values of \(r\); at small \(r\) it goes to zero following the functional derivative (due to steep repulsion that keeps atoms from coming close to one another), while for large distances the discrepancy goes to zero as both potentials tend to zero at the same cutoff. 

\begin{figure} 
  \makebox[\textwidth][c]{\includegraphics[width=6.1in]{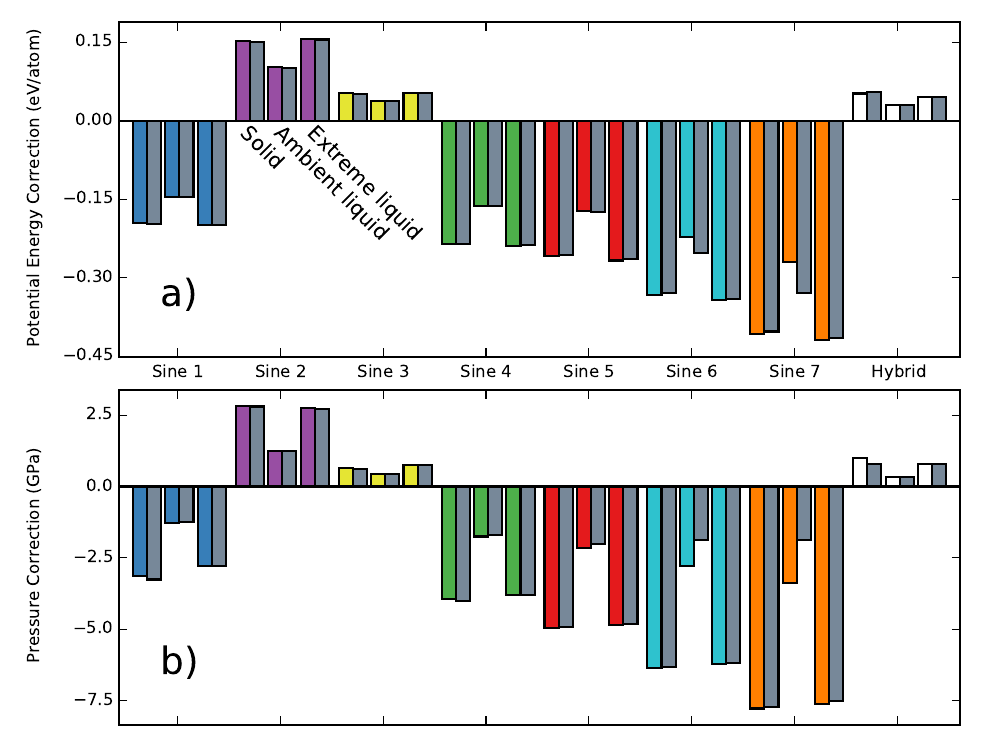}}
  \caption{Comparison of correction from FunUQ (colors matching curves in Fig. \ref{fig:potentials}, with addition of the hybrid potential, shown in white and discussed in sub-section \ref{sec:resultsMorse}) and direct simulation (gray). Each set of three bars shows the solid, ambient liquid, and extreme liquid cases from left to right for each potential. Results for a) potential energy and b) pressure.}
  \label{fig:results}
\end{figure}

The functional error was numerically integrated using the trapezoid rule to obtain the total correction for the QoI. In order to verify these correction predictions we performed explicit MD simulations with the high-fidelity modified LJ potentials. Figure \ref{fig:results} compares the a) potential energy and b) pressure differences explicitly simulated with low and high-fidelity potentials (gray) to the corrections obtained with FunUQ (with colors matching those in Fig. \ref{fig:potentials}, as well as the hybrid LJ-Morse potential discussed in sub-section \ref{sec:resultsMorse} in white). These results are shown in detail in Tables \ref{tbl:PE} and \ref{tbl:P} in the appendix. 

For almost all cases, the FunUQ predictions are in excellent agreement with the direct simulations. Excluding only the Sine 6 and 7 potentials for the ambient liquid (discussed below) average error is 0.600\% and 1.70\% for potential energy and pressure, respectively. We stress that these very accurate corrections are obtained only making use of the simulation with the unmodified LJ potential; no additional MD simulations are required. 

\begin{figure} 
  \makebox[\textwidth][c]{\includegraphics[width=5.45in]{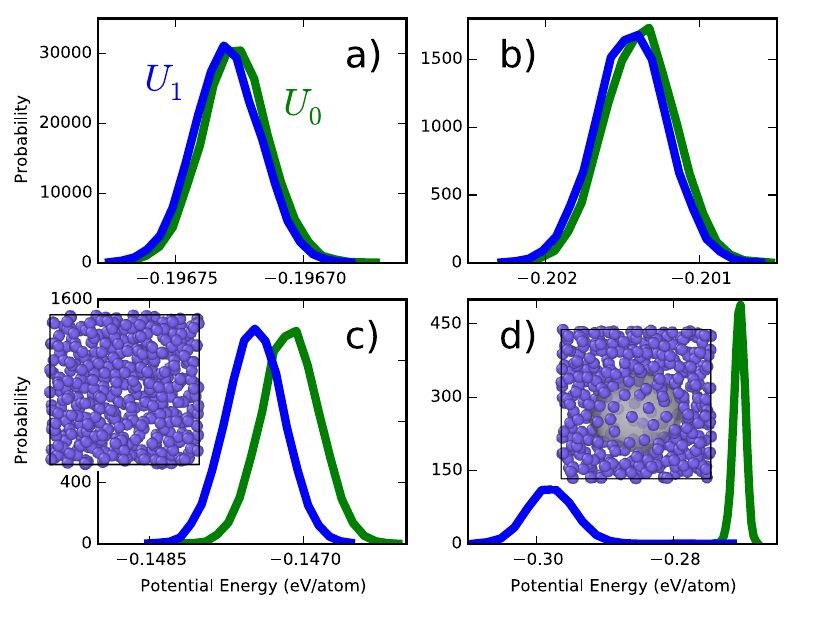}}
  \caption{Examples of potential energy probability distributions showing various degrees of overlap: a) solid with Sine 1, b) extreme liquid with Sine 1, c) ambient liquid with Sine 1, and d) ambient liquid with Sine 7.  Overlap is close to zero for systems with phase change in only one potential; inset d) shows an example of a void formed as compared to the normal liquid in inset c).}
  \label{fig:hist}
\end{figure}

The ability of FunUQ to correct the QoIs is impacted greatly by the degree of phase space overlap between the high and low-fidelity simulations. This can be shown simply by the overlap in histograms of probability distributions of differences in potential energy from the initial to final state, a common practice in free energy calculations \cite{Pohorille2010}. In this case, the states refer to the potential used and the distributions are taken from:
\begin{equation} \label{eq:hist0}
\Delta U_0 = (U(\phi_{HF}, \Gamma_{LF}) - U(\phi_{LF}, \Gamma_{LF}))
\end{equation}
\begin{equation} \label{eq:hist1}
\Delta U_1 = (U(\phi_{LF}, \Gamma_{HF}) - U(\phi_{HF}, \Gamma_{HF}))
\end{equation}
where each term is the energy with potential \(\phi\) and set of samples in phase space \(\Gamma=(\mathbf{x_1}, ..., \mathbf{x_N})\) (each point dependent on the positions of the \(N\) atoms) from the high or low-fidelity potential trajectory.  Examples are shown in Fig. \ref{fig:hist}. More concretely, \(\Delta U_0\) gives the difference between the potential energy distribution from the low-fidelity potential MD simulation and the potential energy distribution from the same atom positions through time re-evaluated with the high-fidelity potential (without additional MD simulation).

Lack of phase space overlap and difference in probability distributions is most significant for the ambient liquid with the Sine 6 and 7 potentials, the same two cases with significantly larger error. Further inspection of the MD trajectory shows that the modified potential results in a structural transformation while the unmodified LJ does not. This is shown in atomistic structures (created using the OVITO software package \cite{Stukowski2010}) included as insets in Fig. \ref{fig:hist} comparing the ambient liquid with Sine 1 and Sine 7:  the liquid undergoes cavitation for Sine 7 (and to a lesser extent with Sine 5 and 6). Thus the low-fidelity LJ explores a vastly different region in phase space as compared to that of the high-fidelity Sine 7; however, even under such unfavorable conditions FunUQ is able to provide some level of correction. In contrast, the solid cases have sharply peaked distributions with strong overlap and correspondingly highly accurate predictions, with the caveat of more difficult convergence of the results. Histograms for all physical cases and potentials are within the supplementary material, Section S3. 

\begin{figure} 
  \makebox[\textwidth][c]{\includegraphics[width=3.3in]{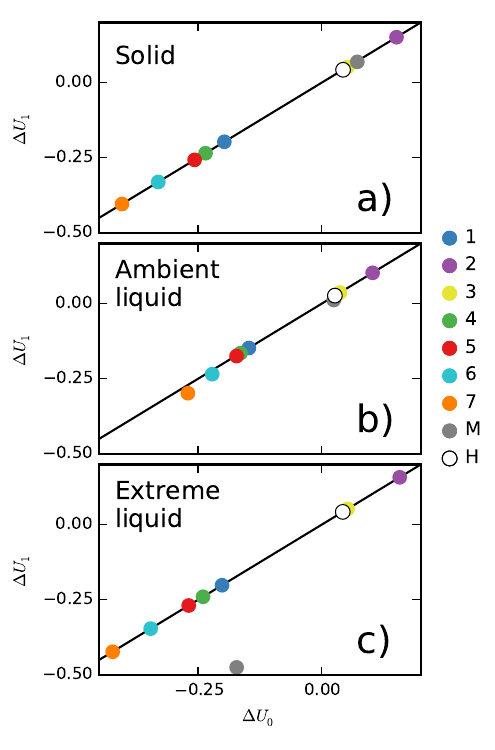}}
  \caption{Similarity in potential energy probability distribution peaks for all potentials (colors matching Figures \ref{fig:potentials} and \ref{fig:results} in sub-section \ref{sec:resultsSine} with the exception of Morse in gray and hybrid LJ-Morse in white, both discussed in sub-section \ref{sec:resultsMorse}). Each physical case separated for readability: a) solid, b) ambient liquid, and c) extreme liquid.}
  \label{fig:histpoints}
\end{figure}

The degree of similarity of phase space exploration can be further summarized by plotting the average of \(\Delta U_0\) against the average of \(\Delta U_1\). The closer to slope of unity, the more significant the distribution overlap. Figure \ref{fig:histpoints} gives results that agree with the histograms in Fig. \ref{fig:hist} and supplementary material, Section S3; namely, for cases where the explored phase space for the low and high-fidelity trajectories is less similar and there is little histogram overlap, the prediction error increases, particularly for the ambient liquid. Results in this figure for the Morse (gray symbols) and hybrid LJ-Morse (white) potentials are discussed in sub-section \ref{sec:resultsMorse}.

\subsection{Correction for the Morse potential and hybrid potential calculations} \label{sec:resultsMorse}
\begin{figure}
  \makebox[\textwidth][c]{\includegraphics[width=3.3in]{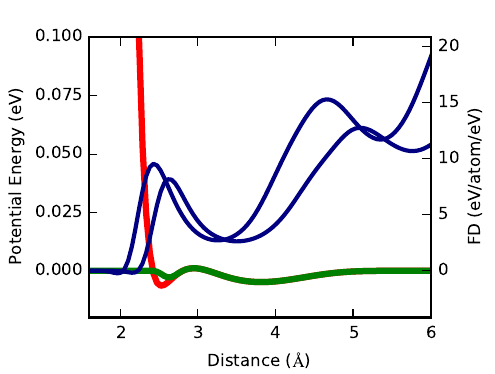}}
  \caption{Discrepancy between LJ and Morse potentials (red) and between hybrid LJ-Morse and Morse (green). Both liquid perturbative functional derivatives shown in blue to demonstrate the non-zero sensitivity at low separation distance.}
  \label{fig:hybrid}
\end{figure}

Predicting properties for the Morse potential from LJ further highlights the challenge of attempting to correct a prediction made with a low-fidelity model. In this case the discrepancy between the two models is very large for short interatomic distances, leading to near zero phase space overlap and an inaccurate FunUQ prediction for all physical cases. The red line in Figure \ref{fig:hybrid} shows this discrepancy as a function of interatomic separation with the functional derivatives from both liquid conditions in blue to demonstrate the sensitivity to changes in the potential at these distances. The short range repulsion is described with an exponential for Morse while LJ uses an inverse power of 12. Thus the discrepancy between the two potentials diverges for short distances and the functional derivative correction, being first order, produces extremely large errors. These results are therefore omitted from Figure \ref{fig:results}. 

Our approach can, nevertheless, be useful in such circumstances, but the high-fidelity model cannot be fully replaced by the low-fidelity one. Instead, we create a hybrid potential in Eq. \ref{eq:hybrid} that smoothly switches from the low-fidelity LJ to the high-fidelity Morse between 2.8 and 2.4\AA\  using error functions, leading to a discrepancy of reasonable magnitude shown in green in Figure \ref{fig:hybrid}.

\begin{equation} \label{eq:hybrid}
\phi_{Hybrid} =
\begin{cases} 
      \phi_{LJ}(r) & r\geq 2.8 \\
      \phi_{LJ}(r)\cdot [B(r)+0.5] + \phi_{Morse}(r)\cdot [-B(r)+0.5] & 2.8 > r > 2.4 \\
      \phi_{Morse}(r) &  2.4 \geq r \\ 
\end{cases}
\end{equation} 
\begin{equation} \label{eq:erf}
     B(r) = 0.5\erf(8(r-2.6))
\end{equation}
With the hybrid potential in Eq. \ref{eq:hybrid} the high-fidelity model is used only where necessary. This is similar in spirit to adaptive sampling methods in multi-scale simulations, with examples in the literature \cite{Knap2008, Barton2011, Roehm2015} and codes available: the Co-design Embedded Visco-Plasticity Proxy Application (CoEVP) \cite{Dorr2014} and the Co-design Heterogeneous Multiscale Method Proxy Application (CoHMM) \cite{RouetLeduc2014}. 

It is then possible to use FunUQ to correct a low-fidelity simulation run with this hybrid potential to the result with the high-fidelity Morse by using the discrepancy between the two potentials and the functional derivative of the hybrid potential. The hybrid potential reduces the discrepancy and the functional approach provides an accurate correction for all cases, as shown with the white bars in Figure \ref{fig:results}. The hybrid potential solid case correction shows an additional example of increased error (5.9\%), again due to poor overlap in explored phase space between the two potentials (see supplementary material, Section S3). Even with these significantly different systems FunUQ provides the majority of the necessary correction. Further, all predictions for Morse from the hybrid potential (white symbols) lie near the unity line in Fig. \ref{fig:histpoints}  (describing good agreement between the phase space of each trajectory) in contrast to Morse from LJ (gray symbols), furthest from the line. We note that the high-fidelity model needs to be used during the actual simulation; however, only approximately 12\% of the atomic force calculations fall in the range requiring the high-fidelity model. Thus, if the high-fidelity model was significantly more intensive than the low-fidelity the methodology would still greatly reduce computational cost.

\section{Discussion}
The perturbative method to obtain the functional derivatives described here is similar to techniques in free energy methods: free energy perturbation and thermodynamic integration \cite{Bash1987, Karplus1990}. One subset of these methods most comparable to FunUQ and often used in biology simulations, referred to as computer alchemy, is utilized to calculate free energy changes along non-physical paths as potentials are slowly turned on or off for the various molecules or solvents of interest \cite{Gumbart2012,He2014,Wescott2002, Lawrenz2012}. 
These methods share with the work described here the need to sample from the phase space trajectory of an initial state (often a simple reference state for free energy calculations, e.g. the Einstein crystal), to ensure that phase space is not too dissimilar to that of the final state of interest, and to converge exponentially weighted averages. However, while free energy methods generally focus on a single path from initial to final state (or a bidirectional path), using FunUQ, once the functional derivative is calculated for a given low-fidelity model, the corrections can be made with respect to any other high-fidelity model as long as minimum conditions are satisfied.
If the low-fidelity simulation does not explore the regions of phase space relevant for the high-fidelity potential at the conditions of interest the results will be poor; this is most striking when switching potentials results in a structural phase transition. A check of  phase space overlap was performed here in order to understand cases where FunUQ is unable to provide accurate error corrections as is done in free energy calculation methods. In real applications where FunUQ would be most useful, i.e. where the high-fidelity function is much more computationally expensive than the low-fidelity function, checking phase space overlap is expensive and should be performed by evaluating the high-fidelity model for relatively few configurations or replacing it with a computationally efficient surrogate model. Additionally, the functional derivative correction can only be used for relatively small functional discrepancies such that its effect can be described to first order. 
We also note that the perturbative approach used here is applicable only to input functions that appear linearly in the Hamiltonian so the perturbation can be separated additively. This is not a general limitation of FunUQ but of the specific approach used here to obtain the functional derivatives. 

This work demonstrates FunUQ in equilibrium simulations. Continuing work should include investigation of functional derivatives in non-equilibrium processes as an analog of the relationships between free energy perturbation and non-equilibrium work methods in free energy calculations. The method can be generally used in equilibrium MD simulations and numerous other physics problems including multi-scale simulations. Ongoing investigations will include using FunUQ to correct predictions in solid mechanics with the plasticity model as the input constitutive law.

An important advantage of FunUQ over uncertainty propagation in parameters is that it enables changes in the actual functional forms used in the simulation. In our example, we are not limited to Lennard-Jones potentials with different parameters but can change the shape of the potential (with the limitations discussed above). This feature makes FunUQ an important tool within continuing UQ research. For example, efforts in quantifying model uncertainty \cite{Zhang2000, Droguett2008, Park2014}
could be expanded to problems with higher computational cost and include an increased number of models by using FunUQ. Our method could also simplify techniques which include both parametric and model UQ \cite{Droguett2013} by utilizing similar forms for each portion of the uncertainty.

In summary we  demonstrated the ability to calculate functional derivatives of a quantity of interest predicted by a non-trivial physics simulation with respect to its input laws. This information can be used to quantify the uncertainties originating in the simulation due to the use of the input function or to correct the prediction if a more accurate model becomes available. We developed a computationally efficient approach to compute the functional derivative in a MD simulations performed using the Lennard-Jones potential and shows that this information can be used to infer thermodynamic properties corresponding to various other potentials without re-running the simulation. The functional uncertainty quantification approach is quite generally applicable and we believe it will be useful to quantify uncertainties in a variety of materials models.

\appendix

\section{Thermodynamic quantities using coordination number}

The calculation of the functional derivative using the perturbative approach (Eq. \ref{eq:FD2} and \ref{eq:perturbative}) is expanded upon here, beginning with the contribution to the Hamiltonian from the perturbation. This quantity can be directly calculated with a sum over all pairs of atoms as the potential energy in any MD simulation:
\begin{equation} \label{eq:PEall}
H' = \sum_{i<j} \phi'(r_{ij}) 
\end{equation}
This would require modification of the MD code to calculate this quantity within the force loop or summing over a saved atomic trajectory. This expression can be replaced:
\begin{equation} \label{eq:PEcoord}
H' = \frac{N}{2} \sum_{k} \phi'(r_k) \cdot c(r_k) 
\end{equation} 
where \(c(r)\) is the average coordination number at a given separation distance \(r\), discretized into \(k\) bins and \(N\) the number of atoms. This is neither invasive to the code, nor requires significant storage or computation past the low-fidelity simulation.

With the potential energy as the QoI, Q is simply the sum of Eq. \ref{eq:PEcoord} and the low-fidelity simulation potential energy. For the pressure we use the virial expression (without the small ideal gas contribution as it is already present in the low-fidelity pressure):
\begin{equation} \label{eq:Pallvec}
P' = \frac{1}{3V}\sum_{i<j} \textbf{f}_{ij} \textbf{r}_{ij} 
\end{equation}
Because we examine cases of two-body central forces this can be simplified:
\begin{equation} \label{eq:Pall}
P' = \frac{1}{3V}\sum_{i<j} f(r_{ij})r_{ij} 
\end{equation}
and with the same motivations as above, we rewrite in terms of the coordination number, again with \(k\) bins in separation distance:
\begin{equation} \label{eq:Pcoord}
P' = \frac{N}{2} \frac{1}{3V}\sum_k f(r_k)\cdot r \cdot c(r_k) 
\end{equation}

This quantity, added to the low-fidelity simulation pressure, provides Q in Eq. \ref{eq:perturbative} for the QoI pressure.

\section{Correction results}
Tables \ref{tbl:PE} and \ref{tbl:P} show full results for all physical conditions for potential energy and pressure, respectively. Figure \ref{fig:results} compares columns \(\Delta Q^{Sim}\) and \(\Delta Q^{FunUQ}\). All results here are calculated with respect to the LJ low-fidelity potential, except the rows marked with (H). Those cases use the low-fidelity hybrid LJ-Morse potential. 

\begin{table}
  \centering
  \caption{Comparison of potential energy correction from functional derivatives and direct simulation.}
  \label{tbl:PE}
  \begin{tabular}{| p{1.15cm} | p{1.85cm} | p{1.3cm} | p{1.3cm} | p{1.55cm} |  p{1.4cm} | p{1.55cm} | p{1.45cm} |}
    \hline
    Temp. & Potential & \(Q_{LF}\) & \(Q_{HF}\) & \(Q_{LF} +\) \(\Delta Q^{FunUQ}\) & \(\Delta Q^{Sim}\) & \(\Delta Q^{FunUQ}\) & \(\Delta Q\) \newline \% Error \\
    \hline
    (K) & -  & \multicolumn{5}{c|}{(eV/atom)} & - \\
    \hline
300 & Sine 1 & -1.08 & -1.28 & -1.27 & -0.197 & -0.196 & 0.456 \\
300 & Sine 2 & -1.08 & -0.928 & -0.927 & 0.151 & 0.152 & 0.784 \\
300 & Sine 3 & -1.08 & -1.03 & -1.03 & 0.0518 & 0.0522 & 0.793 \\
300 & Sine 4 & -1.08 & -1.31 & -1.31 & -0.235 & -0.235 & 0.0132 \\
300 & Sine 5 & -1.08 & -1.33 & -1.34 & -0.256 & -0.259 & 0.988 \\
300 & Sine 6 & -1.08 & -1.41 & -1.41 & -0.330 & -0.333 & 0.993 \\
300 & Sine 7 & -1.08 & -1.48 & -1.49 & -0.403 & -0.407 & 0.995 \\
300 & Morse & -1.08 & -0.998 & -1.02 & 0.0804 & 0.0628 & 21.9 \\
300 & Morse (H) & -1.05 & -0.998 & -1.00 & 0.0552 & 0.0520 & 5.89 \\
1300 & Sine 1 & -0.704 & -0.850 & -0.850 & -0.146 & -0.146 & 0.0158 \\
1300 & Sine 2 & -0.704 & -0.602 & -0.601 & 0.102 & 0.103 & 1.23 \\
1300 & Sine 3 & -0.704 & -0.666 & -0.666 & 0.0377 & 0.0379 & 0.588 \\
1300 & Sine 4 & -0.704 & -0.867 & -0.868 & -0.164 & -0.164 & 0.160 \\
1300 & Sine 5 & -0.704 & -0.878 & -0.876 & -0.174 & -0.172 & 1.08 \\
1300 & Sine 6 & -0.704 & -0.956 & -0.925 & -0.252 & -0.221 & 12.1 \\
1300 & Sine 7 & -0.704 & -1.03 & -0.974 & -0.330 & -0.271 & 18.0 \\
1300 & Morse & -0.704 & -0.655 & -0.662 & 0.0485 & 0.0416 & 14.3 \\
1300 & Morse (H) & -0.703 & -0.673 & -0.673 & 0.0295 & 0.0295 & 0.0295 \\
5000 & Sine 1 & -0.398 & -0.598 & -0.598 & -0.200 & -0.200 & 0.136 \\
5000 & Sine 2 & -0.398 & -0.243 & -0.241 & 0.155 & 0.157 & 0.896 \\
5000 & Sine 3 & -0.398 & -0.346 & -0.345 & 0.0524 & 0.0527 & 0.477 \\
5000 & Sine 4 & -0.398 & -0.636 & -0.637 & -0.238 & -0.239 & 0.441 \\
5000 & Sine 5 & -0.398 & -0.663 & -0.665 & -0.265 & -0.267 & 0.932 \\
5000 & Sine 6 & -0.398 & -0.738 & -0.741 & -0.340 & -0.343 & 0.959 \\
5000 & Sine 7 & -0.398 & -0.814 & -0.818 & -0.416 & -0.420 & 1.01 \\
5000 & Morse & -0.398 & -0.321 & -0.393 & 0.0774 & 0.00547 & 93.1 \\
5000 & Morse (H) & -0.366 & -0.320 & -0.320 & 0.0454 & 0.0455 & 0.376 \\
    \hline
  \end{tabular}
\end{table}

\begin{table}
  \caption{Comparison of pressure correction from functional derivatives and direct simulation.}
  \label{tbl:P}
  \begin{tabular}{| p{1.15cm} | p{1.85cm} | p{1.3cm} | p{1.3cm} | p{1.55cm} |  p{1.4cm} | p{1.55cm} | p{1.45cm} |}
    \hline
    Temp. & Potential & \(Q_{LF}\) & \(Q_{HF}\) & \(Q_{LF} +\) \(\Delta Q^{FunUQ}\) & \(\Delta Q^{Sim}\) & \(\Delta Q^{FunUQ}\) & \(\Delta Q\) \newline \% Error \\
    \hline
    (K) & -  & \multicolumn{5}{c|}{(GPa)} & - \\
    \hline
300 & Sine 1 & -0.116 & -3.37 & -3.25 & -3.26 & -3.13 & 3.89 \\
300 & Sine 2 & -0.116 & 2.69 & 2.70 & 2.81 & 2.81 & 0.279 \\
300 & Sine 3 & -0.116 & 0.514 & 0.538 & 0.630 & 0.654 & 3.75 \\
300 & Sine 4 & -0.116 & -4.12 & -4.05 & -4.01 & -3.94 & 1.73 \\
300 & Sine 5 & -0.116 & -5.03 & -5.07 & -4.91 & -4.95 & 0.763 \\
300 & Sine 6 & -0.116 & -6.43 & -6.48 & -6.32 & -6.37 & 0.792 \\
300 & Sine 7 & -0.116 & -7.83 & -7.90 & -7.72 & -7.78 & 0.805 \\
300 & Morse & -0.116 & 1.43 & -1.31 & 1.54 & -1.19 & 177 \\
300 & Morse (H) & 0.491 & 1.27 & 1.49 & 0.781 & 0.998 & 27.7 \\
1300 & Sine 1 & 0.129 & -1.12 & -1.13 & -1.25 & -1.26 & 1.44 \\
1300 & Sine 2 & 0.129 & 1.37 & 1.36 & 1.24 & 1.23 & 0.498 \\
1300 & Sine 3 & 0.129 & 0.563 & 0.562 & 0.434 & 0.433 & 0.219 \\
1300 & Sine 4 & 0.129 & -1.58 & -1.62 & -1.71 & -1.75 & 2.34 \\
1300 & Sine 5 & 0.129 & -1.87 & -2.03 & -2.00 & -2.16 & 7.70 \\
1300 & Sine 6 & 0.129 & -1.75 & -2.65 & -1.88 & -2.77 & 47.3 \\
1300 & Sine 7 & 0.129 & -1.74 & -3.26 & -1.87 & -3.39 & 81.5 \\
1300 & Morse & 0.129 & 0.0733 & -0.359 & -0.0558 & -0.488 & 774 \\
1300 & Morse (H) & 0.124 & 0.469 & 0.463 & 0.344 & 0.338 & 1.68 \\
5000 & Sine 1 & 54.4 & 51.6 & 51.6 & -2.77 & -2.77 & 0.0187 \\
5000 & Sine 2 & 54.4 & 57.1 & 57.1 & 2.72 & 2.74 & 0.905 \\
5000 & Sine 3 & 54.4 & 55.1 & 55.1 & 0.741 & 0.750 & 1.17 \\
5000 & Sine 4 & 54.4 & 50.6 & 50.6 & -3.80 & -3.81 & 0.161 \\
5000 & Sine 5 & 54.4 & 49.6 & 49.5 & -4.81 & -4.84 & 0.704 \\
5000 & Sine 6 & 54.4 & 48.2 & 48.2 & -6.18 & -6.23 & 0.845 \\
5000 & Sine 7 & 54.4 & 46.9 & 46.8 & -7.53 & -7.61 & 1.12 \\
5000 & Morse & 54.4 & 39.8 & 36.8 & -14.6 & -17.5 & 20.1 \\
5000 & Morse (H) & 39.0 & 39.8 & 39.8 & 0.796 & 0.794 & 0.279 \\
    \hline
  \end{tabular}
\end{table}

\section*{Acknowledgments}
This research was supported by the U.S. Department of Energy (DOE), Office of Advanced Scientific Computing Research (ASCR) through the Exascale Co-Design Center for Materials in Extreme Environments (ExMatEx, exmatex.org), under contract B615461. Computational resources of nanoHUB.org are gratefully acknowledged.

\clearpage

\bibliographystyle{elsarticle-num} 
\bibliography{bibtex/Reeve-Strachan-2016_revisions}

\include{Reeve-Strachan-2016_supplementary}
\texttt{}
\end{document}